\documentclass[acmtog,nonacm]{acmart}

\usepackage{booktabs}
\usepackage[ruled]{algorithm2e}

\SetAlFnt{\small}
\SetAlCapFnt{\small}
\SetAlCapNameFnt{\small}
\SetAlCapHSkip{0pt}

\begin{document}
\title{Joint Optimization of Triangle Mesh, Material, and Light from Neural Fields with Neural Radiance Cache}

\author{Jiakai Sun}
\orcid{0000-0003-2894-9456}
\affiliation{
  \institution{Zhejiang University}
  \country{China}
}

\author{Weijing Zhang}
\affiliation{
  \institution{Zhejiang University}
  \country{China}
}

\author{Zhanjie Zhang}
\affiliation{
  \institution{Zhejiang University}
  \country{China}
}
\author{Guangyuan Li}
\affiliation{
  \institution{Zhejiang University}
  \country{China}
}
\author{Tianyi Chu}
\affiliation{
  \institution{Zhejiang University}
  \country{China}
}
\author{Lei Zhao}
\affiliation{
  \institution{Zhejiang University}
  \country{China}
}
\author{Wei Xing}
\affiliation{
  \institution{Zhejiang University}
  \country{China}
}

\begin{abstract}
Traditional inverse rendering techniques are based on textured meshes, which naturally adapts to modern graphics pipelines,
but costly differentiable multi-bounce Monte Carlo (MC) ray tracing poses challenges for modeling global illumination.
Recently, neural fields have demonstrated impressive
reconstruction quality but fall short in modeling indirect illumination.
In this paper, we introduce a simple yet efficient inverse rendering framework that combines the strengths of both methods.
Specifically, given pre-trained neural field representing the scene,
we can obtain an initial estimate of the signed distance field (SDF) and create a Neural Radiance Cache (NRC),
an enhancement over the traditional radiance cache used in real-time rendering.
By using the former to initialize differentiable marching tetrahedrons (DMTet) and the latter to model indirect illumination,
we can compute the global illumination via single-bounce differentiable MC ray tracing and jointly optimize the geometry,
material, and light through back-propagation. Experiments demonstrate that, compared to previous methods,
our approach effectively prevents indirect illumination effects from being baked into materials,
thus obtaining the high-quality reconstruction of triangle mesh, Physically-Based (PBR) material, and High Dynamic Range (HDR) light probe.
\end{abstract}


%
%

\newcommand {\newtext}[1]{#1}
\newcommand {\mycaption}[1]{#1}
\begin{teaserfigure}
    \centering
      \includegraphics[width=\textwidth]{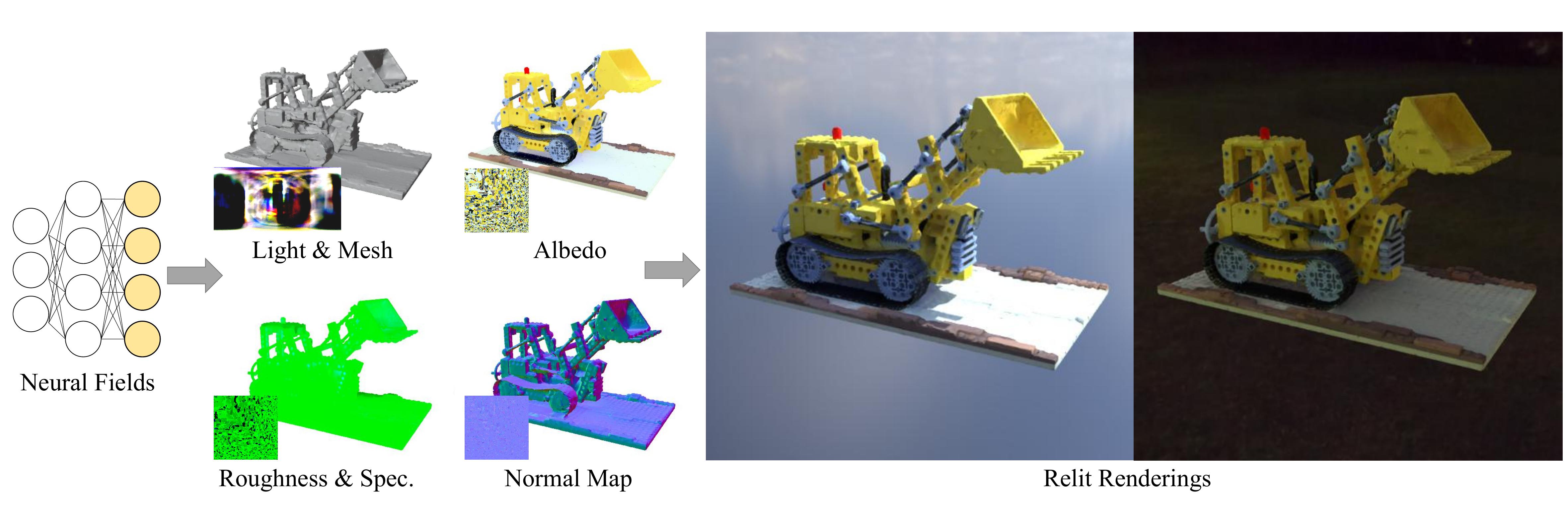}
      \caption{\mycaption{Given the neural field representing a scene, our framework aims to jointly optimize triangle mesh, PBR material, and HDR light probe, which is directly compatible with modern graphics pipelines, facilitating integration with digital content creation tools and game engines.
      The Lego model is from NeRF synthetic dataset (CC-BY-3.0), and the relit renderings are rendered with Blender Cycles and Light probes from Poly Haven (CC0).}}
      \label{fig:teaser}
\end{teaserfigure}

\makeatletter
\let\@authorsaddresses\@empty
\makeatother

\maketitle
\pagestyle{empty}
\section{Introduction}

Coordinate-based neural networks, also known as neural fields~\cite{xie2022neural},
are gaining increasing popularity in the field of computer vision and graphics.
Neural Radiance Fields (NeRF)~\cite{nerf}
and its variants~\cite{barron2021mipnerf,barron2022mipnerf360,verbin2022refnerf, SunSC22, mueller2022instant, Chen2022ECCV, ReluField_sigg_22}
represent radiance fields as neural fields,
leveraging differentiable volume rendering to synthesize photo-realistic novel views.
In addition, surface-based methods~\cite{yariv2020multiview, Niemeyer2020CVPR, wang2021neus, Oechsle2021ICCV, Azinovic_2022_CVPR, yariv2021volume}
model Signed Distance Fields (SDF) using neural fields,
enabling accurate geometric reconstruction.
These neural field-based method have significantly advanced the development of inverse rendering—a longstanding task in computer vision and graphics to estimate physical attributes of a scene from images.
Recent works~\cite{boss2021neuralpil,boss2021nerd,nerfactor,iron-2022,physg2021,zhang2022invrender} utilizing neural fields for geometry, material, and/or light decomposition has yielded impressive results.
However, as these methods are not physically-based,
they can only recover basic illumination effects and tend to bake complex effects such as shadows,
inter-reflections, and color bleeding into materials.
Furthermore, neural fields are not directly compatible with modern graphics pipelines,
which poses limitations on its practical application.

In contrast, textured meshes naturally fit into modern graphics pipelines and can be optimized via physically-based differentiable rendering~\cite{reparam,Path-Space,edgesampling,path_replay},
enabling an unbiased estimation of attributes of a scene.
Nonetheless, multi-bounce differentiable Monte Carlo (MC) rendering poses a steep increase in computational time and memory requirements for gradient-based optimization,
particularly when aiming to jointly optimize geometry, material, and light.
Consequently, previous state-of-the-art methods for joint reconstruction~\cite{Munkberg_2022_CVPR, hasselgren2022nvdiffrecmc} have primarily focused on direct illumination, i.e., environment lighting with shadows.
Although these approaches yield convincing results, they fall short in decoupling material reflectance from indirect illumination effects.

To harness the reconstruction capabilities of neural fields and the efficient performance of modern graphics pipelines,
we propose JOC\footnote{\textbf{J}oint \textbf{O}ptimization of Triangle Mesh, Material, and Light from Neural Fields with Neural Radiance \textbf{C}ache}
(pronounced as "joke"), a framework that integrates cutting-edge techniques from computer vision and graphics. JOC allows for capturing global illumination using single-bounce differentiable MC rendering with a modest increase in cost, and jointly optimizes geometry, material, and light via back-propagation, as shown in Figure~\ref{fig:teaser}.
Specifically, we first obtain a Neural Radiance Cache (NRC)~\cite{Mueller2021NRC}—a low-cost and highly generalizable version of traditional radiance cache—from a trained neural field.
The trained neural field offers an additional advantage: a robust initial estimate of the SDF for the Differentiable Marching Tetrahedrons (DMTet)~\cite{shen2021dmtet}, preventing mismatches between the triangle mesh obtained from DMTet and the NRC.
By integrating the NRC into a highly optimized ray tracing pipeline~\cite{parker2010optix}, we can jointly optimize triangle meshes, Physically-Based (PBR) materials, and High Dynamic Range (HDR) light probe.
Compared to previous inverse rendering methods based on textured meshes, JOC separates indirect illumination effects and material, yielding high quality reconstruction.
Notably, we make no assumptions about the neural fields, other than it being capable of synthesizing novel views and estimating depth via ray casting\footnote{This assumption will be implicit in all subsequent sections for simplicity unless otherwise stated.}. This allows us to train the NRC from the majority of neural field-based methods, facilitating seamless integration with the latest advances in the area of neural fields.

\section{Related Work}

\paragraph{Neural Fields}
In recent years, neural field-based methods demonstrate exceptional capabilities in 3D reconstruction tasks,
attracting widespread attention from researchers in computer vision and graphics.
By parameterizing radiance fields using neural fields and utilizing differentiable volume rendering,
NeRF~\cite{nerf} is able to synthesize photorealistic novel views.
This remarkable performance inspires a series of subsequent works that enhance various aspects of the method,
including reducing training and/or rendering costs~\cite{SunSC22, mueller2022instant, Chen2022ECCV, ReluField_sigg_22,takikawa2022variable,yu_and_fridovichkeil2021plenoxels},
and improving performance in handling sparse inputs~\cite{sun2023vgos,niemeyer2022regnerf,yu2021pixelnerf,Jain_2021_ICCV}
and challenging scenarios~\cite{martin2021nerfw,barron2021mipnerf,barron2022mipnerf360,mildenhall2022nerfd}.
These variants spur the community to create some sophisticated models like Nerfacto~\cite{nerfstudio},
which integrates a variety of advanced techniques.
Some neural implicit surface reconstruction methods~\cite{wang2021neus,Oechsle2021ICCV,yariv2021volume},
exemplified by NeuS~\cite{wang2021neus}, modify the volume rendering formula used by NeRF and use neural fields to parameterize SDF,
achieving high-quality surface reconstructions.
This approach further inspires works that enhance geometric consistency~\cite{fu2022geo}, outdoor scene performance~\cite{sun2022neuconw},
and robustness in sparse settings~\cite{Yu2022MonoSDF}. Inspired by Nerfacto, the community develops NeuS-facto~\cite{Yu2022SDFStudio},
a high-performance NeuS-like model that excels in various aspects.
Despite the powerful reconstruction capabilities of neural field-based methods,
they are not directly applicable to modern graphics pipelines, which limits their practical use.

\paragraph{Inverse Rendering}
Inverse rendering is a longstanding challenge in computer vision and graphics.
Most early works~\cite{dong2014appearance,Xia_Dong_Peers_Tong_2016,gardner2003linear,ghosh2009estimating,guarnera2016brdf}
involve stringent conditions for scene capture, such as the need for specific lighting or complex camera setups.
These extra settings provide sufficient priors.
However, the joint recovery of geometry, material, and light from casually captured image sets under unknown illumination remains a severely ill-posed problem. 

Recently, a series of works~\cite{nerfactor,boss2021nerd,boss2021neuralpil,physg2021,zhang2022invrender} have leveraged the powerful reconstruction capabilities of neural fields to tackle this issue.
NeRFactor~\cite{nerfactor}, NeRD~\cite{boss2021nerd}, and Neural-PIL~\cite{boss2021neuralpil} employ NeRF-like geometry representations, simplified lighting models, and custom rendering pipelines to achieve decomposition of shape and reflectance under unknown Illumination, allowing for view synthesis and relighting.
Notably, among these methods, NeRFactor directly utilizes pre-trained NeRF, while NeRD and Neural-PIL require their own neural field training to meet model requirements.
PhySG~\cite{physg2021}, by utilizing neural fields to parameterize SDF and diffuse albedo, and using spherical Gaussians (SG) to represent environment map and specular bidirectional reflectance distribution function (BRDF), achieves material editing and re-lighting under the assumption of non-spatially-varying reflectance.
InvRender~\cite{zhang2022invrender} trains a visibility multi-layer perceptron (MLP) and an indirect illumination MLP leveraging PhySG. With these MLPs and its custom rendering pipeline, it achieves inverse rendering that accounts for global illumination.

While neural field-based inverse rendering methods exhibit impressive performance,
they require additional processing to be compatible with modern digital content creation (DCC) tools and game engines, which might sacrifices reconstruction quality.
In contrast, when taking textured meshes to represent the scene, these issues do not arise.
NVDIFFREC~\cite{Munkberg_2022_CVPR} utilizes DMTet~\cite{shen2021dmtet} and differentiable rasterizer with deferred shading to jointly optimize mesh, material, and lighting from images.
Yet, due to its simplified shading model, shadows are baked into the material.
NVDIFFRECMC~\cite{hasselgren2022nvdiffrecmc} extends NVDIFFREC with a differentiable MC renderer and a differentiable denoiser, effectively separating shadow and material.
However, considering the increased noise level and drastically increased iteration time caused by multi-bounce path tracing, NVDIFFRECMC only considers direct lighting, leading to coupling of indirect illumination effects with the material.

\paragraph{Radiance Caching}
Caching of radiometric quantities, a technique now prevalent in modern game engines for real-time global illumination, owes its inception to the seminal work of~\cite{Ward:1988:Irradiance} on diffuse interreflection.
Later works~\cite{Greger:1998:Irradiance,Krivanek:2005:Radiance,Ren:2013:rrf,scherzer2012pre, silvennoinen2017real} enhance the performance and/or generalization of caching algorithms.
However, achieving both of these objectives often requires hand-crafted priors and sophisticated data structure.

With the advent of neural rendering, a line of works~\cite{Ren:2013:rrf,Mueller2021NRC,Mueller:2020:ncv,hadadan2021neural} represent the radiance cache with neural networks to circumvent these requirements.
Among these methods, NRC~\cite{Mueller2021NRC} stands out by employing a fully fused MLP for online training, thereby achieving robust real-time radiance caching with predictable performance and resource consumption.
JOC utilizes a streamlined version of NRC, designed to be compatible with most off-the-shelf neural fields techniques.

\paragraph{Concurrent works}
Several concurrent works~\cite{wang2023neural,hadadan2023inverse} also combine neural fields and explicit representations.
We introduce these works and highlight the differences between them and JOC as follows:

\begin{itemize}
  \item FEGR~\cite{wang2023neural} represents the intrinsic properties of the scene using a neural field and estimate the G-buffer with volumetric rendering. However, it cannot model indirect illumination.
  \item InvGI~\cite{hadadan2023inverse} leverages a neural field both to represent the radiance function, and to account for global illumination without building path integrals. However, InvGI does not allow for joint optimization of geometry, material, and light.
\end{itemize}

\begin{figure*}
  \centering
  \includegraphics[width=\textwidth]{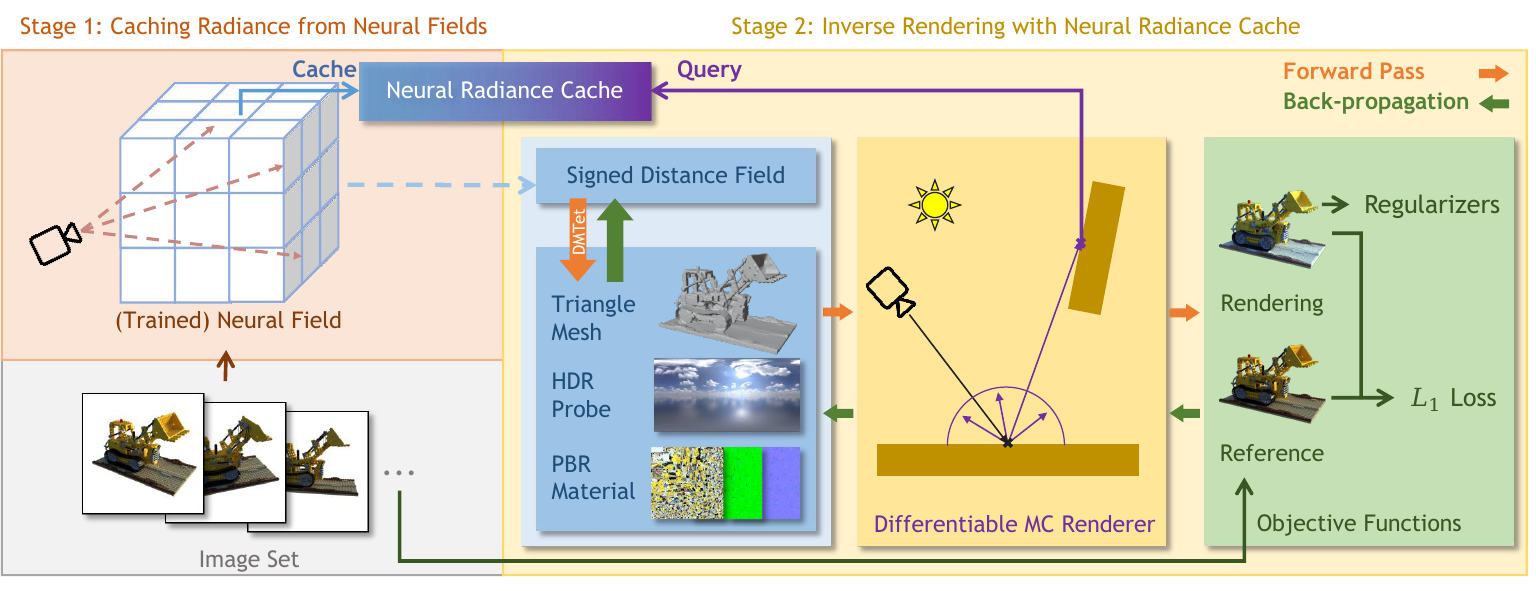}
  \caption{\mycaption{Overview of JOC. Given a set of posed multi-view images and a corresponding neural field trained on it, JOC aims to reconstruct triangle mesh, PBR material and HDR light probe, which can be directly integrated into modern digital content creation tools and game engines.
  More specifically, JOC comprises two stages: In the first stage~(Section~\ref{sec:stg1}),
  the Neural Radiance Cache (NRC)~\cite{Mueller2021NRC} is trained using the neural field, thereby 'caching' the radiance.
  We then jointly optimize the geometry, material, and light utilizing the Differentiable Marching Tetrahedrons (DMTet)~\cite{shen2021dmtet} 
  and a \textbf{single-bounce} differentiable Monte Carlo (MC) renderer combined with NRC in the second stage~(Section~\ref{sec:stg2}).
  Note that we optionally leverage the neural field for a better initial estimate of the Signed Distant Field (SDF) for DMTet.}}
  \label{fig:overview}
\end{figure*}

\section{Background}

As the main idea of JOC involves training NRC from existing Neural Fields, this section will focus on the introduction of Neural Fields in 3D Reconstruction and Neural Radiance Cache.

\subsection{Neural Fields in 3D Reconstruction}
\label{sec:nf3d}
Refer to~\cite{xie2022neural}, we define a Neural Field as a field that is parameterized fully or in part by a neural network which maps position (and optionally, some additional inputs) to corresponding attributes.
For instance, NeRF~\cite{nerf} maps position and viewing direction to density and color, while NeuS~\cite{wang2021neus} maps these to value of the signed distance function and color.
Recent works have demonstrated that neural fields are a powerful scene representation for 3D reconstruction from a set of posed camera images.
Despite the diverse attributes computed by neural fields, the majority of these methods render neural fields to corresponding pixels by casting camera rays, and optimize the parameters of the neural fields through gradient-based optimization to realize 3D reconstruction.
To make this process more formal, we start from the measurement equation~\cite{veach1998robust}
\begin{equation}
  \begin{gathered}
    \label{eqn:measure}
    I=\int_{\mathcal{A} \times \Omega} W_e(x, \omega) L_i(x, \omega) d x d \omega^{\perp},
    \end{gathered}
\end{equation}
where $I$ represents real-valued measurement.
Each measurement corresponds to the integral of the outputs of a sensor that responds to the incident radiance $L_i$  upon it with the sensor responsivity $W_e$ over its area $\mathcal{A}$ and the upper hemisphere $\Omega$.
In the context of rendering images, each measurement $I$ represents the value of a single pixel, while $W_e$ is simplified to tone mapping and gamma correction.
Methods based on neural fields also use the measurement equation to calculate pixel values, but they have different ways to compute $L_i$.
For instance, IDR~\cite{yariv2020multiview} uses surface-based rendering, while NeRF uses volume rendering.
However, they both compute $L_i$ by casting rays and utilizing the attributes computed by the neural fields.
Additionally, most methods based on neural fields can be used to estimate depth, i.e., the distance between the origin of the ray and its intersection point with the scene surface.
For example, the depth can be estimated in NeRF-like model by a similar procedure as rendering RGB, while SDF-based model can directly use sphere tracing to find intersection points.

\subsection{Neural Radiance Cache}
\label{sec:nrc}
In light transport calculations~\cite{veach1998robust}, the incident radiance $L_i$ in Equation~\eqref{eqn:measure} can be obtained using the exitant radiance $L_o$:
\begin{equation}
\label{eqn:lilo}
  L_i(x, \omega)=L_o(x_{\mathcal{M}}(x,\omega), -\omega),
\end{equation}
where $x_{\mathcal{M}}(x,\omega)$ is the ray-casting function, which returns the first point of the scene $\mathcal{M}$ visible from $x$ in direction $\omega$.
Further, we can computes the exitant radiance $L_o$ using the rendering equation~\cite{kajiya1986rendering}~:
\begin{equation}
\label{eqn:render}
  L_o\left(x, \omega\right)=E\left(x, \omega\right)+
  \int_{\Omega} f\left(x, \omega_i, \omega\right) L_i(x, \omega_i) d \omega_i^{\perp}.
\end{equation}
Here, $w$ is the direction of $L_i$, and $f\left(x, \omega_i, \omega\right)$ is the BRDF. As $E\left(x, \omega\right)$ represents the emitted radiance $L_e$,
NRC focuses on the remaining part of the equation, which is used to compute the scattered radiance $L_s$, given that it involves a challenging high-dimensional integral.
To address this issue, NRC approximates $L_s$ with a fully fused MLP expressed as:
\begin{equation}
  NRC_\phi: (x, \omega, n, r, \alpha, \beta) \mapsto L_s,
\end{equation}
caching the the scattered radiance $L_s \in \mathbb{R}^3$ at
position $x \in \mathbb{R}^3$ and scattered direction $\omega \in \mathbb{S}^2$ with surface normal $n \in \mathbb{S}^2$, surface roughness $r \in \mathbb{R}$, 
diffuse reflectance $\alpha \in \mathbb{R}^3$ and specular reflectance $\beta \in \mathbb{R}^3$ as additional parameters. Here we omit the encoding for each parameter for simplicity, please refer to~\cite{Mueller2021NRC} for more details.
Despite employing a minimal set of extra parameters to enhance its robustness, 
original NRC still requires material parameters $r$, $\alpha$, and $\beta$, which are often inaccessible in many neural fields.
Hence, we rely solely on $x$, $\omega$, and $n$ as input parameters,
still yielding satisfactory approximations of indirect illumination.

\section{Method}
Given a set of multi-view images with associated foreground segmentation masks, camera poses, and a pre-trained neural field,
JOC aims to reconstruct triangle meshes, material, and light directly suitable for modern DCC tools and game engines.

As illustrated in Figure~\ref{fig:overview}, our framework consists of two stages:
The first stage~(Section~\ref{sec:stg1}) trains the NRC using the existing neural field.
As NRC was designed for online training in real-time rendering, the traning time of this stage primarily depends on the rendering speed of the neural field.
The second stage~(Section~\ref{sec:stg2}) jointly optimizes the triangle mesh, material, and light through DMTet and single-bounce differentiable MC rendering combined with NRC.

\begin{figure}
  \includegraphics[width=\linewidth]{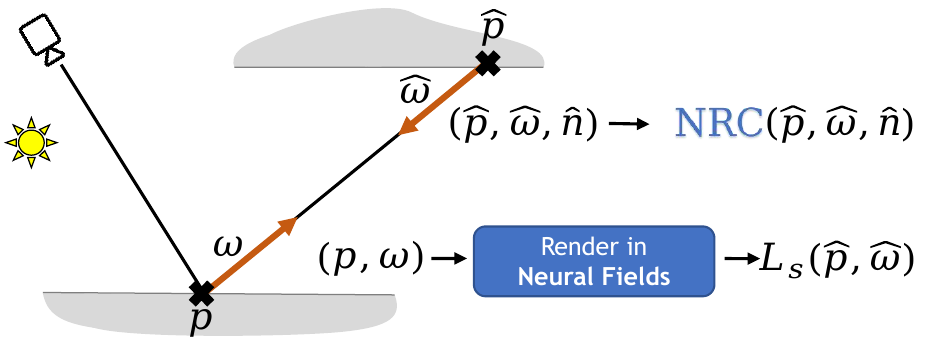}
  \caption{\mycaption{Visualization of caching radiance. As neural fields are able to render RGB images and estimate depth, we can utilize them to capture scattered radiance at surface points, subsequently 'cache' it into the NRC.}}
  \label{fig:caching}
\end{figure}

\subsection{Caching Radiance from Neural Fields}
\label{sec:stg1}
As introduced in Section~\ref{sec:nrc},
NRC is a MLP that takes position, scattered direction, and surface normal as inputs and outputs the exitant radiance of position $x$ in scattered direction $\omega$.
Following instant-ngp~\cite{mueller2022instant}~, we use a multi-resolution hash grid~\cite{mueller2022instant}~to encode positions and one-blob encoding~\cite{Muller2019NIS}~for the remaining input parameters.

Since the neural field can compute RGB value and estimate depth by casting rays, we can treat it as a function that
\begin{equation}
\label{eqn:nfrt}
  F_\theta: (o,\omega) \mapsto (L_i, p).
\end{equation}
Here, $o\in\mathbb{R}^3$ is the ray origin, $\omega \in \mathbb{S}^2$ is the ray direction,
$L_i \in \mathbb{R}^3$ is the incident radiance at $(o,\omega)$, and $p=x_{\mathcal{M}}(o,\omega) \in \mathbb{R}^3$ is the intersection point of the ray and the scene $\mathcal{M}$.
According to Equation~\eqref{eqn:lilo}, the incident radiance $L_i$ at $(o,\omega)$ is equivalent to the exitant radiance $L_o$ at $(p,-\omega)$.
By replacing $o$ in Equation~\eqref{eqn:nfrt} with surface points $p$, we can obtain the exitant radiance $L_o$ from $-\omega$, which is formulated as:
\begin{equation}
\label{eqn:surface_scatter}
  L_o(\hat{p}, -\omega), \hat{p} = F_\theta(p,\omega),
\end{equation}
where $\hat{p}=x_{\mathcal{M}}(p,\omega)$.
Assuming non-emitted objects, we get scattered radiance $L_s=L_o$ from surfaces from Equation~\eqref{eqn:surface_scatter}, which can be used to train NRC.
To be more specific, following InvRender~\cite{zhang2022invrender},
we randomly sample camera rays from a pre-trained neural field $F_\theta$,
estimate the depth to get surface points $p$, and then sample rays for each surface point.
If these rays intersect with the surface, we record the intersection position $\hat{p}$, scattered direction $\hat{\omega}=-\omega$, and compute the analytic surface normal $\hat{n}$ of the intersection point for each intersection;
see Figure~\ref{fig:caching} for the visualized process.
Finally, we use the luminance-relative $\mathcal{L}_2$ loss~\cite{Mueller2021NRC} to train the NRC, which is formulated as:
\begin{equation}
\label{eqn:nrcloss}
  \mathcal{L}_2\left(L_s(\hat{p}, \hat{\omega}), NRC_\phi(\hat{p}, \hat{\omega}, \hat{n})\right)
  =\frac{\left(L_s(\hat{p}, \hat{\omega})-NRC_\phi(\hat{p}, \hat{\omega}, \hat{n})\right)^2}{(NRC_\phi(\hat{p}, \hat{\omega}, \hat{n}))^2+0.01},
\end{equation}
where the gradient of the normalized factor is not back-propagated in optimization.
\subsection{Inverse Rendering with Neural Radiance Cache}
\label{sec:stg2}
As a proof of concept, we extend NVDIFFRECMC~\cite{hasselgren2022nvdiffrecmc}, which achieves joint optimization of triangle mesh, material, and light by employing a differentiable MC renderer that only consider direct illumination and a differentiable denoiser.
In contrast, JOC enhance the renderer using NRC to account for global illumination, achieving superior reconstruction quality compared to NVDIFFRECMC.
\subsubsection{Representations of Scene Parameters}
\paragraph{Geometry}
We use SDF to represent geometry and leverage DMTet to convert it into a triangle mesh. SDF-based neural fields naturally provide an initial value for DMTet, while recent work~\cite{lin2023magic3d} converts the density field to an SDF by subtracting a non-zero constant,
allowing NeRF-like neural fields to also provide initial guess for the SDF. It should be noted that we can optimize geometry from scratch without relying on any initial estimate.
\paragraph{Material}
We represent materials using Disney's PBR BRDF model and normal maps. These are widely used in modern game engines to enhance realism and scene details in physically-based shading models.
\paragraph{Light}
We use an HDR light probe to represent illumination. This allows us to directly relight the reconstructed objects in modern renderers.
\subsubsection{Global Illumination with a Single Bounce}
In the realm of physically-based rendering, the integral in the rendering equation are most often estimated using Monte Carlo sampling and ray tracing, as expressed by the following equation:
\begin{equation}
  \begin{aligned}
  L_s\left(x, \omega\right) & =\int_{\Omega} f\left(x, \omega_i, \omega\right) L_i(x, \omega_i) d \omega_i^{\perp} \\
  & \approx \frac{1}{N} \sum_{j=1}^N \frac{f\left(x, \omega_j, \omega\right) L_i\left(x, \omega_j\right)(\omega_j \cdot n)}{p\left(\omega_j\right)},
  \end{aligned}
\end{equation}
with $N$ samples drawn from some distribution $p\left(\omega_j\right)$.
Moreover, Multiple Importance Sampling (MIS)~\cite{veach1995optimally} is adopted, which is a prevalent sampling technique to reduce variance in Monte Carlo integration and is formulated as:
\begin{equation}
  \sum_{i=1}^n \frac{1}{n_i} \sum_{j=1}^{n_i} w_i\left(X_{i, j}\right) \frac{g\left(X_{i, j}\right)}{p_i\left(X_{i, j}\right)}, \quad w_i(x)=\frac{n_i p_i(x)}{\sum_k n_k p_k(x)}.
\end{equation}
Here, $p_i$ are differnet sampling distributions , $g(x)$ is the integrand and $w_i$ are the weights.
Please refer to~\cite{hasselgren2022nvdiffrecmc} for further details about the sampling strategy.
In NVDIFFRECMC, as multi-bounce ray tracing comes with increased cost and noise, rays only bounce once, that is, only shadow rays are traced.
In this case, computing $L_i$ simplifies to
\begin{equation}
\label{eqn:directLighting}
  L_i(p,\omega)=\begin{cases}\mathbf{E}(p,\omega) &\text{if $\omega$ is not blocked by $\mathcal{M}$}\\0 &\text{otherwise}\end{cases}
\end{equation}
where $\mathbf{E}(p,\omega)$ is the direct lighting estimation from the HDR light probe.
However, if $\omega$ is blocked, $L_i(p,\omega)$ should represent the scattered radiance at the intersection point on the surface of $\mathcal{M}$.
This shading model is clearly biased, resulting in indirect illumination being erroneously baked into the material.
Nevertheless, the NRC serves as an effective solution for modeling indirect illumination.
By modifying Equation~\eqref{eqn:directLighting} as:
\begin{equation}
  L_i(p,\omega)=\begin{cases}\mathbf{E}(p,\omega) &\text{if $\omega$ is not blocked by $\mathcal{M}$}\\NRC_\phi(\hat{p}, -\omega, \hat{n}) &\text{otherwise}\end{cases} 
\end{equation}
where $\hat{p}$ is the intersection point and $\hat{n}$ is the surface normal.
we are able to calculate global illumination via a single bounce, with an acceptable increase in computational cost.

\begin{table*}
  \centering
  \caption{Relighting results for NeRFactor dataset with per-scene quality metrics.
  The presented metrics are the averge over eight different view points under eight different light probes.
  Note that in the settings of experiments on Drums and Ficus, we disabled depth peeling for both JOC and NVDIFFRECMC~\cite{hasselgren2022nvdiffrecmc} due to GPU memory issues,
  which is different from the released code of NVDIFFRECMC.}
  \begin{tabular}{l|ccc|ccc|ccc|ccc|c}
      \toprule
      {} & {} & Hotdog & {} & {} & Lego & {} & {} & Drums$^\star$ & {} & {} & Ficus$^\star$ & {} & {}\\
      {} & PSNR$\uparrow$ & SSIM$\uparrow$ & LPIPS$\downarrow$ & PSNR$\uparrow$ & SSIM$\uparrow$ & LPIPS$\downarrow$ & PSNR$\uparrow$ & SSIM$\uparrow$ & LPIPS$\downarrow$ & PSNR$\uparrow$ & SSIM$\uparrow$ & LPIPS$\downarrow$ \\
      \cmidrule{2-13}
      NeRFactor & 25.452 & 0.912 & 0.124 & 20.325 & 0.832 & 0.140 & 21.570 & 0.908 & 0.086 & 21.636 & 0.917 & 0.095   \\
      NVDIFFREC & \underbar{29.460} & 0.933 & \underbar{0.091} & 19.483 & 0.818 & \underbar{0.135} & 22.388 & 0.912 & \textbf{0.083} & \textbf{27.831} & \textbf{0.961} & \textbf{0.048}   \\
      NVDIFFRECMC & 28.606 & \underbar{0.941} & 0.101 & \underbar{20.464} & \underbar{0.835} & 0.145 & \underbar{22.750} & \underbar{0.917} & 0.086 & 25.162 & 0.942 & 0.060   \\
      JOC     & \textbf{30.688} & \textbf{0.948} & \textbf{0.080} & \textbf{22.065} & \textbf{0.854} & \textbf{0.128} & \textbf{22.949} & \textbf{0.919} & \underbar{0.084} & \underbar{25.905} & \underbar{0.948} & \underbar{0.056} \\

      \bottomrule
  \end{tabular}
  \label{tab:nerfactor}
\end{table*}

\begin{table}
  \caption{Relighting results for NeRF synthetic dataset with per-scene quality metrics.
  Specifically, we choose lego and hotdog, the most commonly used scenes in the dataset.
  The presented metrics are the averge over eight different view points under eight different light probes. }
  \resizebox{1.0\columnwidth}{!}{
  \begin{tabular}{l|ccc|ccc|c}
      \toprule
      {} & {} & Hotdog & {} & {} & Lego & {} & {} \\
      {} & PSNR$\uparrow$ & SSIM$\uparrow$ & LPIPS$\downarrow$ & PSNR$\uparrow$ & SSIM$\uparrow$ & LPIPS$\downarrow$\\
      \cmidrule{2-7}
      NVDIFFRECMC & 25.516 & 0.926 & 0.124 & 22.359 & 0.865 & 0.126 & \\
      JOC     & \textbf{26.822} & \textbf{0.931} & \textbf{0.106} & \textbf{23.024} & \textbf{0.881} & \textbf{0.108} \\
      \bottomrule
  \end{tabular}
  }
  \label{tab:nerf}
\end{table}

\begin{table*}
  \centering
  \caption{Ablation study on relighting results for NeRFactor dataset with per-scene quality metrics.
  The presented metrics are the averge over eight different view points under eight different light probes.}
  \begin{tabular}{l|ccc|ccc|ccc|ccc|c}
      \toprule
      {} & {} & Hotdog & {} & {} & Lego & {} & {} & Drums & {} & {} & Ficus & {} & {}\\
      {} & PSNR$\uparrow$ & SSIM$\uparrow$ & LPIPS$\downarrow$ & PSNR$\uparrow$ & SSIM$\uparrow$ & LPIPS$\downarrow$ & PSNR$\uparrow$ & SSIM$\uparrow$ & LPIPS$\downarrow$ & PSNR$\uparrow$ & SSIM$\uparrow$ & LPIPS$\downarrow$ \\
      \cmidrule{2-13}
      JOC~\textit{w/o SDF Init.} & 29.570 & 0.941 & 0.092 & \textbf{22.091} & 0.853 & 0.128 & \textbf{23.001} & \textbf{0.920} & \textbf{0.082} & 25.338 & 0.944 & 0.059 \\
      JOC     & \textbf{30.688} & \textbf{0.948} & \textbf{0.080} & 22.065 & \textbf{0.854} & 0.128 & 22.949 & 0.919 & 0.084 & \textbf{25.905} & \textbf{0.948} & \textbf{0.056} \\
      \bottomrule
  \end{tabular}
  \label{tab:initsdf}
\end{table*}

\subsubsection{Optimization}
Following~\cite{hasselgren2022nvdiffrecmc}, we jointly optimize the geometry, material and light by minimizing the loss:
\begin{equation}
  \mathcal{L}=\mathcal{L}_{\mathrm{image}}+\lambda_{\mathbf{k}_\mathrm{d}}\mathcal{L}_{\mathbf{k}_\mathrm{d}}+\lambda_{\mathbf{k}_{\mathrm{orm}}}\mathcal{L}_{\mathbf{k}_{\mathrm{orm}}}+\lambda_{{n}}\mathcal{L}_{{n}}+\lambda_{{n}^{\prime}}\mathcal{L}_{{n}^{\prime}}+\lambda_{\mathrm{light}}\mathcal{L}_{\mathrm{light}},
\end{equation}
where $\mathcal{L}_{\mathrm{image}}$ is simply the $\mathcal{L}_1$ loss between the tone mapped rendering and the input image $\mathbf{I}_\text{GT}$, and $\lambda$s are corresponding weights for each regularizers. 

$\mathcal{L}_{\mathbf{k}_\mathrm{d}}$ ,$\mathcal{L}_{\mathbf{k}_{\mathrm{orm}}}$ and $\mathcal{L}_{{n}}$ are the smoothness regularizers for albedo $\mathbf{k}_{\mathrm{d}}$, specular parameters $\mathbf{k}_{\mathrm{orm}}$ and surface normal $n$ with the template as:
\begin{equation}
\mathcal{L}_{P}=\frac{1}{\left|\mathrm{x}_{\text {surf }}\right|} \sum_{x \in \mathbf{x}_{\text {surf }}}\left|P(x)-P(x+\boldsymbol{\epsilon})\right|,
\end{equation}
where $P$ is  $\mathbf{k}_{\mathrm{d}}$, $\mathbf{k}_{\mathrm{orm}}$ or $n$, $x_{surf}$ denotes a set of surface points and $\boldsymbol{\epsilon} \sim \mathcal{N}(0, \sigma=0.01)$ is a small random displacement vector.

$\mathcal{L}_{{n}^{\prime}}$ enforce the tangent space normal map close to $(0,0,1)$ to prevent the coupling between the normal map and the environment light, which is formulated as:
\begin{equation}
  \mathcal{L}_{{n}^{\prime}}=\frac{1}{\left|\mathrm{x}_{\text {surf }}\right|} \sum_{x \in \mathbf{x}_{\text {surf }}}1-\frac{n^{\prime}\left(x\right)+n^{\prime}\left(x+\boldsymbol{\epsilon}\right)}{\left|n^{\prime}\left(x\right)+n^{\prime}\left(x+\boldsymbol{\epsilon}\right)\right|} \cdot(0,0,1).
\end{equation}

$\mathcal{L}_{\mathrm{light}}$ is the monochrome image loss between the demodulated lighting terms and the reference image to disentangle material parameters and light,which is formulated as:
\begin{equation}
\mathcal{L}_{\mathrm{light}}=|Y(\mathbf{c}_d+\mathbf{c}_s)-V(I_\text{GT})|,
\end{equation}
where $Y(x)$ is the luminance operator, $V(x)$ computes the value component of HSV, $\mathbf{c}_d$ and $\mathbf{c}_s$ are diffuse and specular lighting, separately. 

\begin{figure}
  \includegraphics[width=\linewidth]{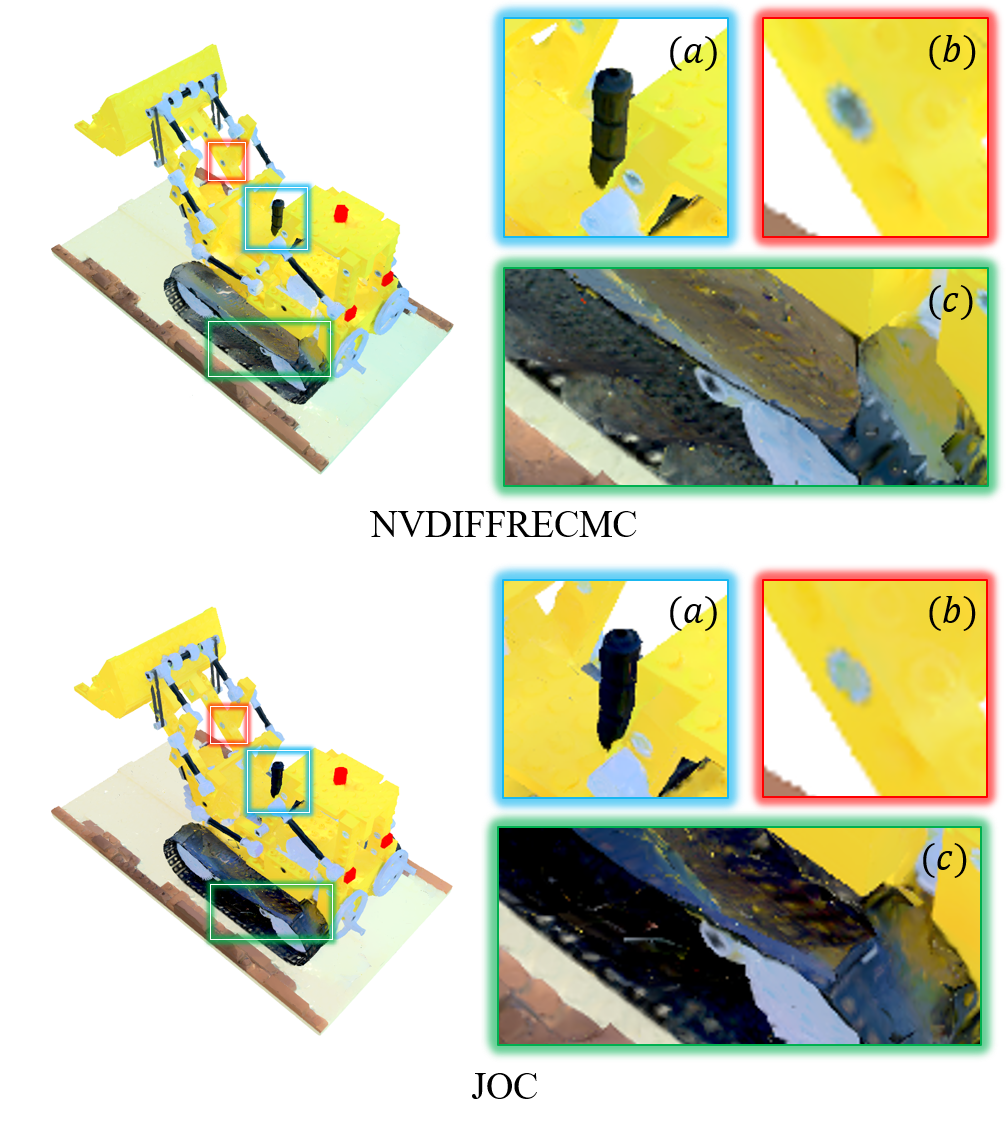}
  \caption{Comparsion between reconstructed albedo of JOC and NVDIFFRECMC~\cite{hasselgren2022nvdiffrecmc} from Lego scene in NeRF synthetic dataset.
  Thanks to modeling indirect illumination in inverse rendering,
  JOC successfully prevents baking (a) inter-reflections (b) holes and (c) color bleeding into albedo.}
  \label{fig:matsep}
\end{figure}

\section{Evaluation Studies}
\begin{figure*}
  \centering
  \includegraphics[width=0.7\textwidth]{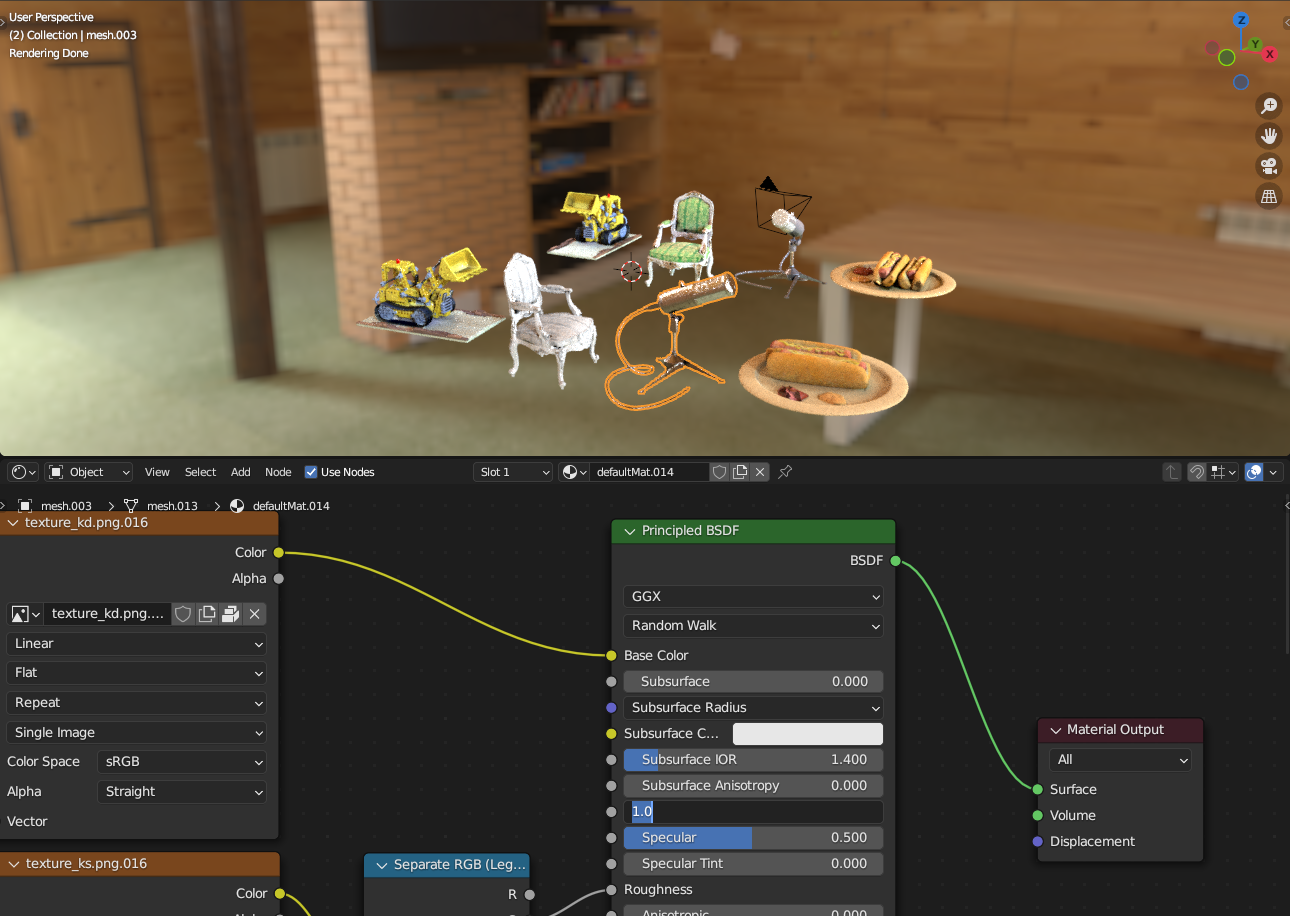}
  \caption{\mycaption{We can use the reconstruction results in modern DCC tools , i.e., Blender, for various applications as scene editing and material editing.}}
  \label{fig:blender}
\end{figure*}

To the best of our knowledge, only a few works achieve joint optimization of triangle mesh, PBR material texture, and light under unknown illumination.
Therefore, we use NVDIFFRECMC~\cite{hasselgren2022nvdiffrecmc}, NVDIFFREC~\cite{Munkberg_2022_CVPR}, and NeRFactor~\cite{nerfactor} as our baselines.
The first two approaches share the same task as ours, while NeRFactor, as a representative of neural field-based methods, is taken as a baseline in their works.
We compare the quality of reconstruction with baselines in Section~\ref{sec:cpr},
and conduct ablation study to show the effect of leveraging better SDF initial estimate from neural fields in Section~\ref{sec:abl}.
Besides, we show the efficiency of NRC in Section~\ref{sec:nrcexp}. 
Our reconstructions are directly suitable for modern digital content creation tools and game engines, as shown in Figure~\ref{fig:blender}.

Note that InvRender~\cite{zhang2022invrender} also leverage off-the-shelf neural scene representation methods as priors,
however, the reconstruction results of InvRender are implicit, which are unsuitable for modern graphics pipelines to render.
Despite this, following InvRender, we utilize IDR~\cite{yariv2020multiview} as the neural fields in all presented experiment results.
\subsection{Comparsions}
\label{sec:cpr}
\begin{figure*}
  \centering
  \includegraphics[width=0.92\textwidth]{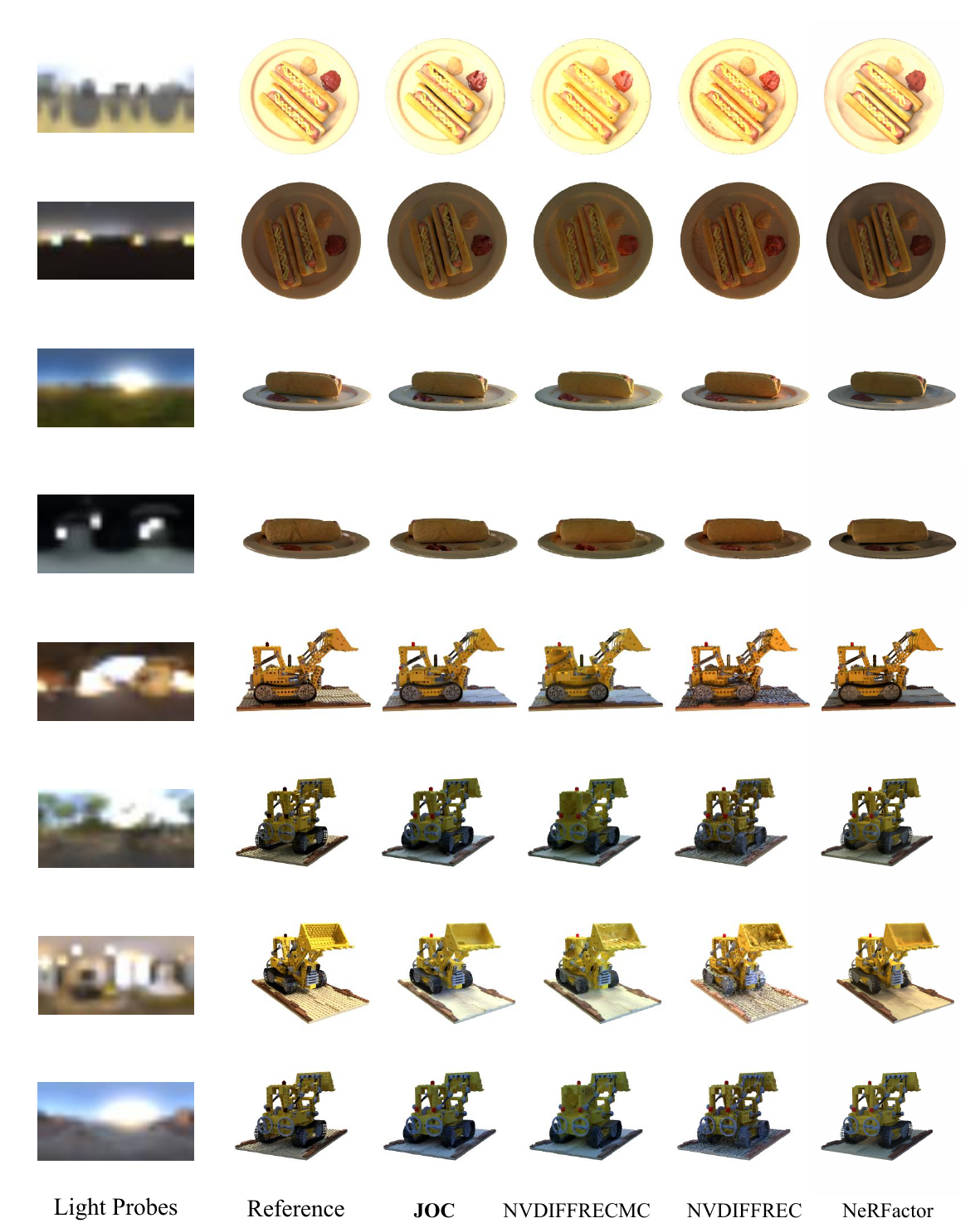}
  \caption{\mycaption{Qualitative comparisons on the NeRFactor~\cite{nerfactor} synthetic dataset. Please zoom-in for details.}}
  \label{fig:viscpr}
\end{figure*}

\begin{figure*}
  \centering
  \includegraphics[width=0.92\textwidth]{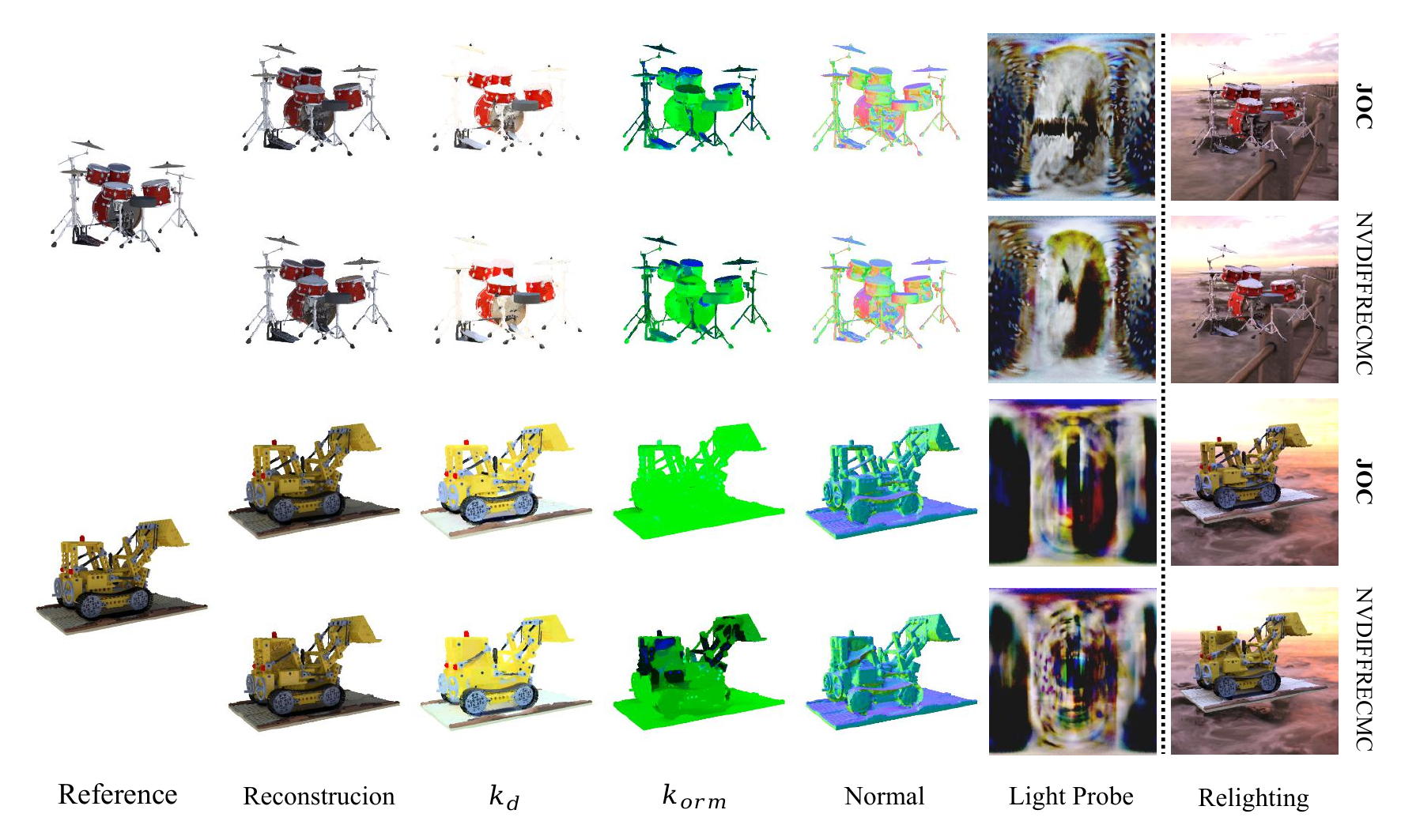}
  \caption{\mycaption{Reconstruction examples from the NeRFactor~\cite{nerfactor} synthetic dataset and relit renderings under a light probe from Poly Haven (CC0).
  Our results outperform the previous work~\cite{hasselgren2022nvdiffrecmc} in terms of visual quality, surface reconstruction and material separation. Please zoom-in for details.}}
  \label{fig:vismat}
\end{figure*}

Table~\ref{tab:nerfactor} presents the quantitative results on the NeRFactor~\cite{nerfactor} synthetic dataset from individual scenes in terms of PSNR, SSIM, and LPIPS.
Furthermore, we provide qualitative comparison in Figure~\ref{fig:viscpr} for relighting results and Figure~\ref{fig:vismat} for reconstruction quality.
Additionally, as the training images in NeRFactor dataset are lit under low frequency lighting,
we conduct extra experiments with training images in NeRF~\cite{nerf} synthetic dataset which contains high frequency lighting and show the results in Table~\ref{tab:nerf}.
Following previous works~\cite{hasselgren2022nvdiffrecmc, Munkberg_2022_CVPR, nerfactor},
we normalize the relit renderings by the average intensity of the reference. 
Our model achieves competitive performance in both quantitative and qualitative comparisons.

Figure~\ref{fig:matsep} illustrates the advantage of modeling indirect illumination in inverse rendering, namely, 
avoiding the baking (a) inter-reflections (b) holes and (c) color bleeding into the material.
\subsection{Ablation Study}
\label{sec:abl}
Table~\ref{tab:initsdf} shows the effect of leveraging the neural field to initialize the SDF for DMTet in our experiments.
For simple scene like Hotdog and Ficus,
the better initial estimate leads to better performance.
However, the surface prior limits the search space for geometric reconstruction,
thus slightly degrading the reconstruction quality for complex scene as Drums and Lego.

\begin{figure}
  \includegraphics[width=\linewidth]{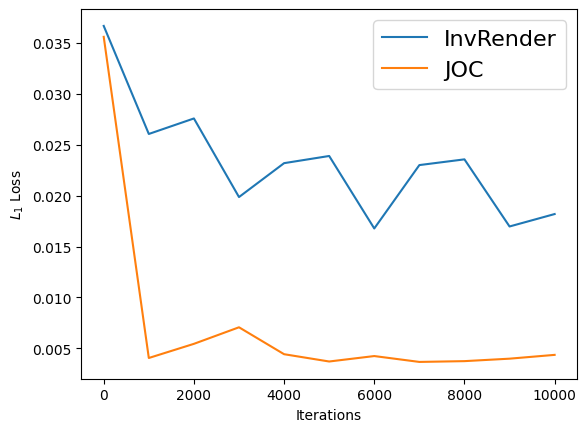}
  \caption{Comparing the training processes of InvRender and JOC (NRC) for representing indirect illumination:
  InvRender predicts the indirect illumination received by surface points and represents it with SGs,
  training by supervising the predicted incoming radiance with the corresponding outgoing radiance.
  In contrast, JOC directly predicts the scattered radiance at surface points via NRC.
  For a fair comparison, JOC also employs $L_1$ loss as the objective function in the experiment depicted.
  It is evident that JOC achieves lower loss within fewer iteration steps.
  Both models exhibit similar iteration time (41 minutes versus 38 minutes for 10,000 iterations)}.
  \label{fig:ilcpr}
\end{figure}

\begin{figure}
  \includegraphics[width=\linewidth]{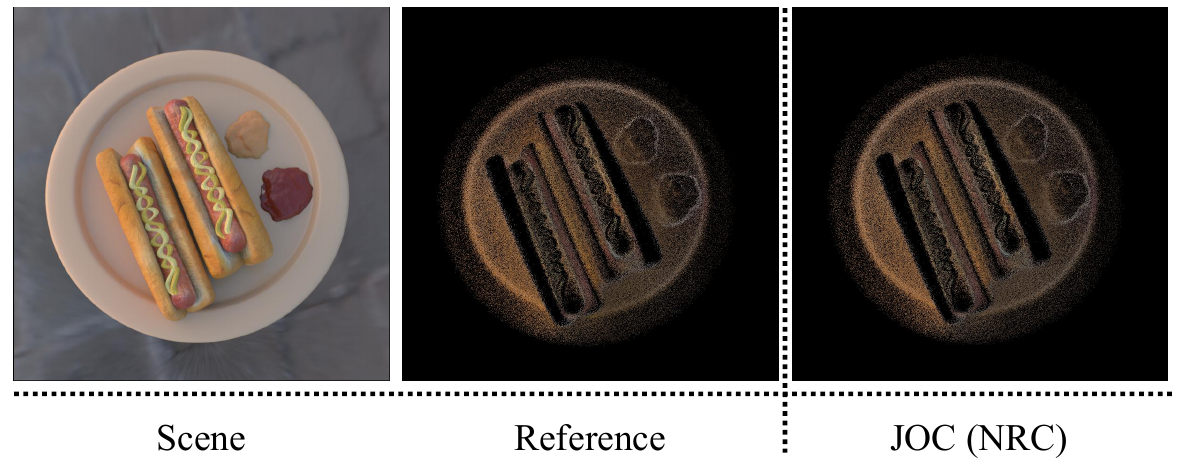}
  \caption{The ability of NRC to cache the scattered radiance. NRC faithfully reproduces the scattered radiance captured from IDR~\cite{yariv2020multiview}}
  \label{fig:ilvis}
\end{figure}

\subsection{Efficiency of Neural Radiance Cache}
\label{sec:nrcexp}
\paragraph{Caching}

InvRender~\cite{zhang2022invrender} is another work that utilizes a neural network to model indirect illumination.
To be specific, InvRender takes a pre-trained IDR~\cite{yariv2020multiview} as the outgoing radiance field, and trains a network that maps a 3D location to its indirect incoming illumination represented as a mixture of SGs.
In contrast, NRC directly caches the exitant radiance of a 3D location with the scattered direction and the analytic normal.
Figure~\ref{fig:ilcpr} illustrates the training of NRC compared to InvRender on the Hotdog scene in the dataset provided by InvRender.
Besides, Figure~\ref{fig:ilvis} shows the ability of NRC to cache the scattered radiance.

\paragraph{Querying}
We modify the renderer of NVDIFFRECMC~\cite{hasselgren2022nvdiffrecmc} which is based on Optix7~\cite{parker2010optix} for utilizing the NRC.
To be specific, we record all requisite information after each sampling and pass it to the NRC within Python,
resulting in additional overheads such as Optix start-up costs, thread synchronization, and data I/O. Despite our implementation,
as a proof of concept, not being highly optimized,
these costs remain manageable:
With 128 samples per pixel (spp), the iteration time in comparison to NVDIFFRECMC has increased by 40\%,
while demanding an additional 20\% of GPU memory.
Yet, to model indirect illumination in a differentiable MC renderer,
these overheads are deemed justifiable,
considering the prohibitive time and memory consumption associated with conventional
multi-bounce differentiable MC renderer.
For instance, in a conventional differentiable Monte Carlo renderer,
the storage requirement grows linearly with number of bounces and quickly exhausts all available GPU memory.
In contrast, we directly simulate infinite-bounce transport via NRC~\cite{Mueller2021NRC}.
For a more comprehensive understanding of the costs associated with multi-bounce differentiable MC rendering, please refer to \cite{path_replay}.

\section{Discussion}
Although the encoding techniques we utilize for the NRC slightly relax the need for fine geometry~\cite{mueller2022instant,Muller2019NIS},
a neural field capable of representing relatively accurate geometry is still necessary.
Furthermore, while NVDIFFRECMC~\cite{hasselgren2022nvdiffrecmc} supports the reconstruction of transparency objects, our approach has not yet been extended to handle such objects.
Future work could potentially explore enhancing the robustness of the NRC suitable for neural fields and expanding support for a wider variety of objects.

\section{Conclusion}
We propose a novel framework that reconstructs triangle meshes,
Physically-Based materials,
and High Dynamic Range light probes under unknown illumination.
Compared to previous methods, we effectively model indirect illumination in modern graphics pipelines through the Neural Radiance Cache~\cite{Mueller2021NRC} trained on neural fields with minimal assumptions,
thereby enhancing the separation of illumination effects and reconstructed materials.
Qualitative and quantitative evaluations demonstrate the effectiveness of our method.
\clearpage
\clearpage
\clearpage
\bibliographystyle{ACM-Reference-Format}
\bibliography{inv}

\end{document}